\documentclass[reprint,amssymb,superscriptaddress]{revtex4-1}

\def\bs{\text{bs}}
\def\E{\text{env}}

\def\er{\text{err}}
\def\P{\text{pr}}
\def\nP{n_\text{pr}}
\def\nE{n_\text{env}}
\def\q{\text{q}}
\def\cl{\text{c}}
\def\conj#1{#1^*}
\def\onenorm#1{\left\lVert#1\right\rVert_1}
\def\ar{r}

\usepackage{amsfonts}
\usepackage{dsfont}
\usepackage{bbold}
\usepackage{bm}
\usepackage{times}
\usepackage{verbatim}
\usepackage[utf8]{inputenc}
\usepackage[T1]{fontenc}
\usepackage[normalem]{ulem}
\usepackage[bottom]{footmisc}
\usepackage{amsmath}
\usepackage{amssymb}
\usepackage{mathrsfs}
\usepackage{mathtools}
\usepackage{amsthm}
\usepackage{physics}

\makeatletter
\def\maketag@@@#1{\hbox{\m@th\normalfont\normalsize#1}} 
\makeatother

\usepackage{graphicx}
\usepackage{float}
\usepackage{subfigure} 
\usepackage{dcolumn}

\usepackage[dvipsnames]{xcolor}
\definecolor{blueprl}{RGB}{46,48,146}
\definecolor{bluemath}{RGB}{94,129,181}
\definecolor{redmath}{RGB}{235,98,53}
\definecolor{greenmath}{RGB}{143,176,50}
\definecolor{yellowmath}{RGB}{255,191,1}
\usepackage[colorlinks=true,urlcolor=blueprl,citecolor=blueprl,linkcolor=blueprl]{hyperref}

\usepackage{pgfplots}
\usepgfplotslibrary{groupplots}

\newtheorem{lemma}{Lemma}
\newtheorem{theorem}{Theorem}
\newenvironment{prooft1}[1][Proof (Theorem 1)]{\noindent\textbf{#1.} }{\ \rule{0.5em}{0.5em}}
\newenvironment{prooft2}[1][Proof (Theorem 2)]{\noindent\textbf{#1.} }{\ \rule{0.5em}{0.5em}}
\newenvironment{proofb}[1][Proof]{\noindent\textbf{#1.} }{\ \rule{0.5em}{0.5em}}

\def\ANU{Centre for Quantum Computation and Communication Technology, Department of Quantum Science, Australian National University, Canberra, ACT 2601, Australia.}
\def\NUS{Centre for Quantum Technologies, National University of Singapore,
3 Science Drive 2, Singapore, Republic of Singapore}
\def\NTU{School of Physical and Mathematical Sciences, Nanyang Technological University, Singapore 637371, Republic of Singapore}

\begin{document}

\title{Optimal probes for continuous variable quantum illumination}
\author{Mark Bradshaw}
\affiliation{\ANU}
\author{Lorc\'{a}n O. Conlon}
\email{lorcan.conlon@anu.edu.au}
\affiliation{\ANU}
\author{Spyros Tserkis}
\affiliation{\ANU}
\author{Mile Gu}
\email{ceptryn@gmail.com}
\affiliation{\NTU}
\affiliation{\NUS}
\affiliation{Complexity Institute, Nanyang Technological University, Singapore 637335}
\author{Ping Koy Lam}
\affiliation{\ANU}
\affiliation{\NTU}
\author{Syed M. Assad}
\email{cqtsma@gmail.com}
\affiliation{\ANU}
\affiliation{\NTU}

\begin{abstract}
Quantum illumination is the task of determining the presence of an object in a noisy environment. We determine the optimal continuous variable states for quantum illumination in the limit of zero object reflectivity. We prove that the optimal single mode state is a coherent state, while the optimal two mode state is the two-mode squeezed-vacuum state. We find that these probes are not optimal at non-zero reflectivity, but remain near optimal. This demonstrates the viability of the continuous variable platform for an experimentally accessible, near optimal quantum illumination implementation.
\end{abstract}
\maketitle

\section{Introduction} Quantum illumination (QI) was introduced by \citet{Lloyd_S_08} for discrete-variable states, showing how entangled photonic probes can be utilized to determine the presence of a weakly reflecting object in a region filled with background noise. This scheme was extended to continuous-variable states by \citet{Tan_etal_PRL_08}, who showed that two-mode Gaussian  entangled states outperform single-mode coherent ones. The advantage that entanglement offers in this task has been also discussed in various other works \cite{Shapiro_Lloyd_NJP_09, Zhuang_Zhang_Shapiro_PRL_17, Zhang_etal_PRA_14, Nair_Gu_arxiv_20, LasHeras_SR_17, Shapiro_PRA_09}, which has in turn inspired a great deal of experimental research \cite{Zhang_etal_PRL_15, Lopaeva_etal_PRL_13, England_Balaji_Sussman_PRA_19, Aguilar_etal_PRA_19, Zhang_etal_PRL_13}. Of particular interest is recent research into microwave QI \cite{Barzanjeh_etal_PRL_15, Xiong_etal_AP_17, Barzanjeh_etal_SA_20, Luong_etal_IEEE_19, Luong_etal_IEEE_18}, as at microwave frequencies background radiation naturally contains many photons which are not present at optical frequencies. This is because the bright thermal background is the region where the advantages of QI are more pronounced. 

In order to fully exploit the benefits quantum mechanics has to offer
for illumination it is essential to know which states are optimal.  In
the discrete-variable case, the optimal probe states to minimise the
probability of error for QI are known~\cite{Yung_Meng_Zhao_arxiv_18}, but the same problem for the
continuous-variable case has remained unsolved. In this paper, we
investigate the optimal probe for QI in the limit of zero object
reflectivity.  We prove that the coherent state and two-mode squeezed
vacuum (TMSV) state are the optimal probes depending on whether
entanglement is allowed or not.  We also show that for all
reflectivities, the optimal probe in the two mode case always has the
same form, a Schmidt decomposition in the Fock basis. This greatly
simplifies the task of numerically finding the optimal probe state.
We find that the numerically optimized non-Gaussian state offer a very limited
 advantage over Gaussian states. Because Gaussian
states can be easily generated in laboratories, these results
suggest that producing the optimal probe for QI is much easier in
continuous-variable compared to discrete-variable systems.

\section{Problem formulation} 
\begin{figure}[t]
\includegraphics[width=0.9\columnwidth]{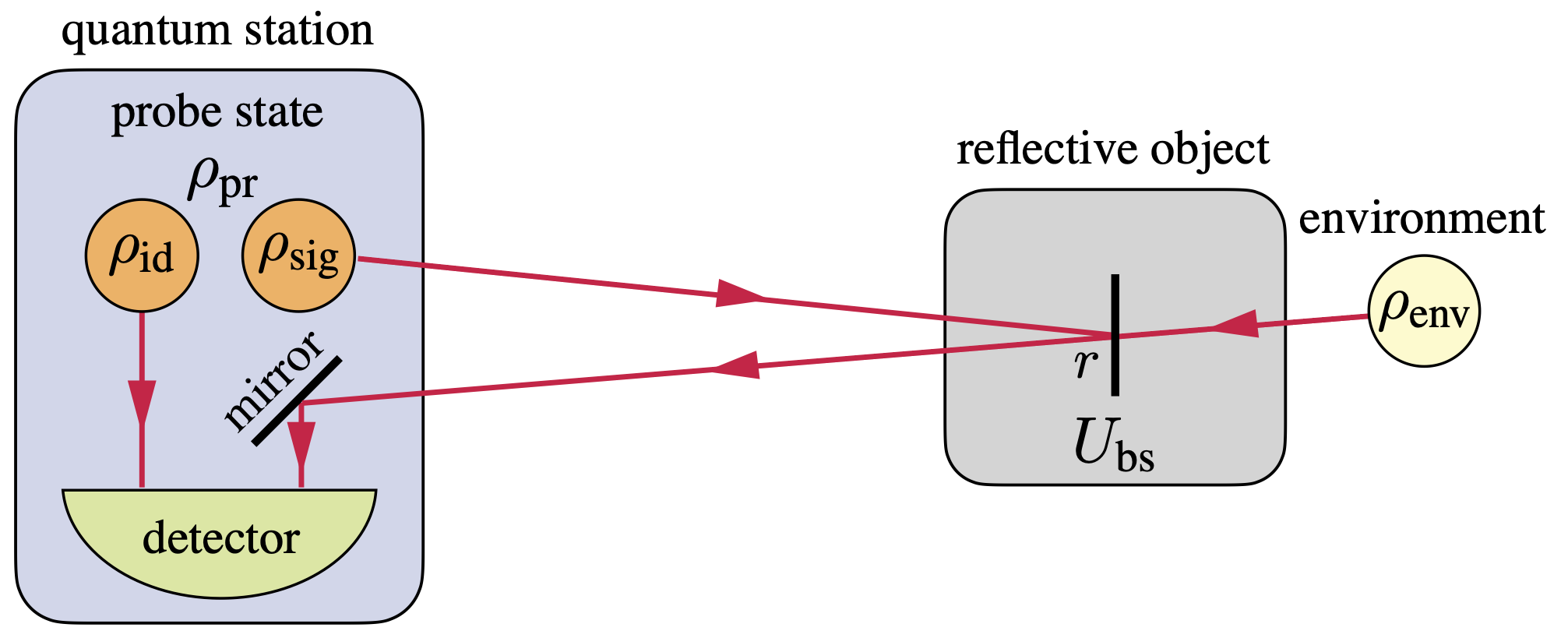}
\caption{{\bf Schematic for quantum illumination.} 
   The partially reflective object, modelled as a beam-splitter with
  reflectivity $r$, is located in a noisy environment represented as a
  state $\rho_\E$ entering the beam-splitter. The signal state
  $\rho_{\text{sig}}$ mixes with the noisy environment and is reflected
  to the detector. For two-mode QI the signal $\rho_{\text{sig}}$ and
  idler $\rho_{\text{id}}$ are entangled and for single-mode QI they
  are not.}
\label{fig_scheme}
\end{figure}

QI is the task where a probe state
$\rho_\P$ is used to detect the presence of a reflective object,
masked by a noisy environment $\rho_\E$, by performing measurements on
the received state. We represent the environment as a state diagonal in the Fock basis
\begin{equation}
\label{eq_env}
\rho_\E =  \sum_m \lambda_m \ketbra{m} \,.
\end{equation} 
We consider two cases: (i) $\rho_\P$ is a
single-mode state, and (ii) $\rho_\P$ is a two-mode state. For the
case of a two-mode probe, we call one mode the {\em signal} which is
sent to the region which may contain the object, and the other mode
the {\em idler} which is reserved for detection purposes (see
Fig.~\ref{fig_scheme}). The reflective object is modelled as 
a beam-splitter of reflectivity $r$, through which the environment couples into the system.
 When the object is present the signal
interacts with the object through the beam-splitter operation $U_\bs(\ar)$
and the received state is then 
\begin{equation}
\label{eq_env}
\rho_1 = \tr_\E \left[ U_\bs(r) (\rho_\P \otimes \rho_\E) U_\bs^{\dag}(r) \right] \,.
\end{equation}
Here, 
\begin{equation}
U_\bs(r) = \exp \left[\theta(r)(\hat{a}^\dagger\hat{b}-\hat{a}\hat{b}^\dagger) \right]\,
\label{beamsplitter}
\end{equation}
is the beam-splitter operator with reflection coefficient $r=\text{sin }\theta$, $\hat{a}$ and $\hat{b}$ are the annihilation
operators for the signal and environment mode, and $\tr_\E$ denotes
partial trace over the mode that is lost to the environment. 
When the object is absent, the reflection coefficient $r$ is set to zero and the
received state is $\rho_0=\rho_\text{idler} \otimes \rho_\E$. Our task is
to discriminate between the two states $\rho_0$ and $\rho_1$ as
accurately as possible. We assume that we have no prior knowledge
about the presence or absence of the object. We do not place any restrictions on the
  detector and only consider the theoretical bounds that can be
  achieved using the best possible receiver.

Suppose a single measurement is made to determine what the received state is, then the minimum error probability in distinguishing the two states is given by the Helstrom bound~\cite{Helstrom_JSP_69}
\begin{equation} p_\er =
    \frac{1-  \frac{1}{2}\onenorm{\rho_0 - \rho_1 }}{2}\,,
\end{equation}
where $\onenorm{A}$ is the trace norm of the matrix $A$, 
and we have assumed a uniform prior.  If instead, we have $M$ copies of the
probe state $\rho_\P$, the probability of error for discriminating
between $\rho_0^{\otimes M}$ and $\rho_1^{\otimes M}$ for large $M$ is given by the quantum Chernoff bound~\cite{Audenaert_etal_PRL_07, Nussbaum_Szkola_AS_09}
\begin{equation}
\label{kapeqn}
p_\er(M) \approx \kappa^M \,, \quad \text{with} \quad \kappa =
\min_{0\leqslant s \leqslant 1} \tr \left( \rho_{1}^s \rho_{0}^{1-s}\right) \,.
\end{equation}
Our aim is to find the optimal probe state for both single-mode and two-mode QI that minimizes the quantum Chernoff bound subject to the constraint that the mean photon number of the probe equals to $\nP$. This constraint is necessary to avoid probe states with unbounded energy.

Both the Helstrom bound and the quantum Chernoff bound are concave
functions of $\rho_\P$, and the domain in which we optimize over is a
convex space because density matrices form a convex set. Thus, based
on Bauer's maximum principle \cite{Bauer_ADM_58}, the optimal probes
are obtained at extremal points, i.e., pure states. Unfortunately,
this problem is in general quite challenging ~\cite{Pardalos_Vavasis_JGO_91,
  Sahni_JC_74}. However, in the region of interest for QI---the low
reflectivity limit---we are able to simplify the problem and obtain
analytic results.  This limit is particularly relevant because this is
where QI promises to be most beneficial. This is also the operating
regime for microwave illumination where most of the signal is lost.

\section{Low reflectivity approximation} When the object reflectivity $\ar$ is low, the state detected is similar to the environment leading to a high probability of error. In this case, we can consider the approximation
\begin{equation}
\rho_{1} \approx \rho_{0} + \ar \Delta \rho \,,
\label{approx}
\end{equation}
where $\Delta\rho = \partial \rho_1/\partial \ar$ evaluated at $\ar=0$. Under this approximation, the Helstrom bound becomes
\begin{equation}
p_\er \approx \frac{1- \ar \,\onenorm{ \Delta\rho} }{2} \,.
\end{equation}
The quantum Chernoff bound for discriminating the two
states $\rho_0$ and
$\rho_{1} = \rho_{0} + \ar\Delta\rho$ for small $\ar$ is $\kappa^M$
with
\begin{equation}
  \label{lemma_qc_approx}
  \kappa \approx 1 - \frac{r^2}{2} \sum_{jk} \frac{\left|\bra{\phi_j}\Delta\rho\ket{\phi_k}\right|^2 }{(\sqrt{\lambda_j}+\sqrt{\lambda_k})^2} \,,
\end{equation}
where $\ket{\phi_j}$ are the eigenvectors of $\rho_0$ with
corresponding eigenvalues $\lambda_j$~\cite{Audenaert_etal_PRL_07}.
To calculate $\Delta\rho$, we approximate the beam-splitter operation [see Eq.~(\ref{beamsplitter})] for a small reflectivity, using the approximation $\theta \approx \ar$, i.e.,
\begin{equation}
U_\bs(r) \approx 1+ \ar(\hat{a}^\dagger\hat{b}-\hat{a}\hat{b}^\dagger) \,.
\end{equation}
Thus, for small $\ar$, we have
\begin{multline}
\label{eq_bs_approx}
U_\bs(r)\ket{n,m}\approx\Big(\ket{n,m} +\ket{n+1,m-1} \ar\sqrt{(n+1)m} \\
-\ket{n-1,m+1} \ar\sqrt{n(m+1)}\Big) \,,
\end{multline}
where $\ket{n,m}$ represents the Fock state with $n$ and $m$ photons in each mode.

\section{Optimal single-mode QI} With this approximation, and a pure probe state $\rho_{\text{pr}}=\ketbra{\psi}$ with
\begin{equation}
  \ket{\psi} = \sum_n c_n \ket{n}\,,  
\end{equation}
the derivative $\Delta\rho$ becomes
\begin{multline}
  \Delta\rho=\sum_m
  \sqrt{m+1}\left(\lambda_{m+1}-\lambda_{m}\right)\\
\times \big(\gamma
  \ketbra{m+1}{m} + \gamma^* \ketbra{m}{m+1} \big) \,,
\end{multline}
where
\begin{equation}
\gamma = \sum_n c_{n+1}c_n^* \sqrt{n+1} \,.
\end{equation}
Since $\Delta\rho$ only depends on the probe state through $\gamma$, to find the optimal probe state we just have to find the state that maximizes $|\gamma|$. The same probe will minimize both the single measurement and multiple-measurement error probabilities.  The optimization result is stated in the following theorem.

\begin{theorem}
\label{th_coherent}
The single mode state subject to an energy constraint
$\nP$ that maximizes  $|\gamma|$ is the coherent state $\ket{\alpha}=\sum_n c_n \ket{n}$, where
\begin{equation}
\label{eq_coherenteq}
c_n = e^{-|\alpha|^2/2} \frac{\alpha^n}{\sqrt{n!}} \,,
\end{equation}
with $\alpha=\sqrt{\nP}$. This is also the probe state
that minimizes the Helstrom bound and quantum Chernoff bound in the
limit of zero object reflectivity.
\end{theorem}

\begin{prooft1}
  Since $\gamma= \sum_n c_{n+1}\conj{c}_n \sqrt{n+1} \leqslant \sum_n |c_{n+1}||\conj{c}_n|\sqrt{n+1}$, we can restrict ourselves to the case in which all $c_n$ are real. The task is then to maximize
  \begin{equation}
    \gamma = \sum_n c_{n+1} c_n \sqrt{n+1}\,,    
  \end{equation}
  subject to  the normalization and energy constraints
  \begin{equation}
    \sum_n c_n^2 =1\,,\text{ and }\;  \sum_n n\, c_n^2 =\nP\,.
  \end{equation}
  Using the method of Lagrange multipliers the necessary condition for the maximum is
  \begin{equation}
    c_{n+1} \sqrt{n+1} +c_{n-1} \sqrt{n} +2 c_n (\mu_1 + n \mu_2)=0 \,,
  \end{equation}
  where $\mu_1$ and $\mu_2$ are the two Lagrange multipliers. Next, we
  show that the coherent state satisfies these set of
  equations. Substituting in Eq.~\eqref{eq_coherenteq}, we get
    \begin{gather}
      \frac{\alpha^{n+1}\sqrt{n+1}}{\sqrt{(n+1)!}} +
      \frac{\alpha^{n-1} \sqrt{n}}{\sqrt{(n-1)!}} +\frac{2\alpha^n (\mu_1 + n \mu_2)}{\sqrt{n!}}=0 \nonumber \\
      \Rightarrow		\alpha +\frac{n}{\alpha}+2(\mu_1+n \mu_2)=0 \,,
    \end{gather}
  which is satisfied by choosing $\mu_1=-\alpha/2$ and $\mu_2=-1/2\alpha$. Hence a coherent state is a solution, and all that is left to do is choose $\alpha=\sqrt{\nP}$ such that it satisfies the energy constraint.
\end{prooft1}

We note that since $\gamma$ does not depend on the environment, the coherent state will be optimal for any environment that is diagonal in the Fock basis. In particular, when the environment is a thermal state with mean photon number $\nE$, so that the coefficients $\lambda_m$ in Eq.~\eqref{eq_env} reads $\lambda_m=\nE^m/(1+\nE)^{m+1}$, the quantum Chernoff bound for single-mode QI with a coherent state probe has
\begin{equation}
  \kappa_\cl=1-\frac{r^2 \nP \left( 1+\nE\right)}{\left[ 1+\nE +\sqrt{\nE(1+\nE)}\right]^2}\, ,
\end{equation}
which is consistent with Ref.~\cite{Tan_etal_PRL_08}.

\section{Optimal two-mode QI} Let us now consider the case where the probe $\rho_{\text{pr}}$ is a two-mode state. The optimal probe in this case can be expressed in the form (see Supplementary material \cite{supplmat} for details)
\begin{equation}
\ket{\psi} = \sum_n c_n \ket{n,n} \,,
\end{equation}
which is the Schmidt decomposition in the Fock basis (see also
Ref.~\cite{Sharma_etal_NJP_18}). The fact that the optimal probe state
can always be written in this form, regardless of object reflectivity, greatly simplifies the task of numerically finding the optimal probe. Describing the environment in the Fock basis, in the same way as for single-mode QI, we detect the following state when the object is absent
\begin{equation}
\label{eq_qi_rho1}
\rho_{0} = \sum_{m} \lambda_m \ketbra{m} \otimes\sum_{n} c_n^2 \ketbra{n} \,.
\end{equation}
Utilizing the approximations in Eqs.~\eqref{approx} and
\eqref{eq_bs_approx}, $\Delta\rho $ in two-mode QI becomes
\begin{multline}
\label{eq_qi_drho}
\Delta\rho = \sum_{nm} \sqrt{m+1}\sqrt{n+1} (\lambda_{m+1}-\lambda_{m})c_n c_{n+1}  \\
{\times}\Big(\ketbra{m+1,n+1}{m,n} + \ketbra{m,n}{m+1,n+1}\Big).
\end{multline}
Unlike the single-mode QI case, $\Delta\rho$ cannot be written as a
sum involving the environment multiplied by a sum involving the
probe. Thus, in general, the optimal probe for two-mode QI depends on
the environment. In what follows, we restrict to the case with a
thermal environment. In the limit of infinite copies of the state, the
TMSV state is the optimal state that minimizes the error
probability. This result is summarized in the following theorem.
\begin{theorem}
  \label{th_epr}
  In the limit of zero object reflectivity, the two-mode probe for QI in
  a thermal environment that maximizes the quantum Chernoff bound
  subject to a constraint on the probe mean photon number of
  $\nP$ is the TMSV state
  $\ket{\psi} = \sum_n c_n \ket{n,n}$ where
  $c_n
  =\sqrt{\nP^n/(\nP+1)^{n+1}}$.
\end{theorem}
\begin{figure}
\includegraphics[width=\columnwidth]{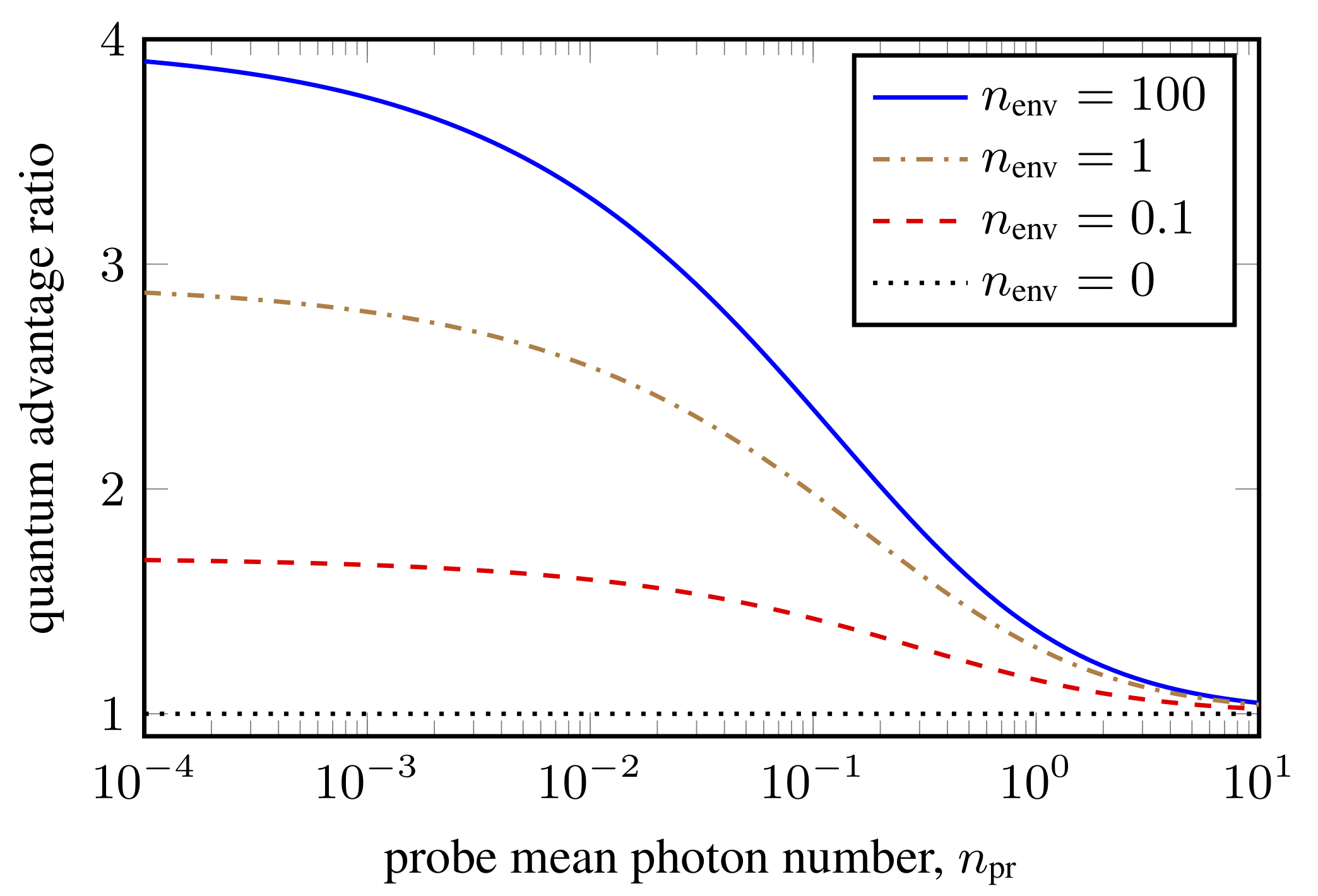}
\caption{{\bf Performance comparison between the optimal two-mode and
    single-mode probes for QI in the limit of zero object reflectivity.} The quantum
  advantage ratio determines the multiple of additional copies
  required for the single-mode probe to have the same error
  performance as the two-mode probe. This has a maximum value of $4$
  when the thermal environment has a high mean photon number
  $\nE$. When $\nE=0$, there is no advantage to using a two-mode probe.
 }
\label{fig_qiadvantage}
\end{figure}
To prove theorem \ref{th_epr}, we require the following lemma, (see Lemma 1 of the Supplemental Material  \cite{supplmat} for prood). 
\begin{lemma}
\label{lemma_epr_mean}
Let $G$ be the function
\begin{equation}
\label{eq_meanmax}
G=\sum_n (n+1) \mathcal{M}(c_n^2, k c_{n+1}^2) \,,
\end{equation} 
where $\mathcal{M}(x,y)$ is a mean function~\footnote{A {\em mean
    function} is a function that defines an `average value' of two
  numbers. Familiar examples of mean functions are the arithmetic
  mean, harmonic mean and geometric mean. A function
  $\mathcal{M}(x,y)$ is a mean function if it can be written as
  $\mathcal{M}(x,y) = y f({x}/{y})$ where $f$ is: (i) monotone
  increasing, (ii) $x f^{-1}(x)=f(x)$ so that $\mathcal{M}$ is
  symmetric, and (iii) $f(1)=1$. See for
  example Refs.~\cite{petz1996monotone, hiai2009riemannian}.}  and $k$ is a positive
constant. Then, with the constraints
  \begin{equation}
    \sum_n c_n^2 =1\,,\text{ and }\;  \sum_n n\, c_n^2 =\bar{n}\,,
  \end{equation}
  $G$ is locally maximized when
  $c_n^2 =\bar{n}^n/(\bar{n}+1)^{n+1}$. If $\mathcal{M}(x,y)$
  is concave it is also globally optimal.
\end{lemma}

\begin{prooft2}
  From Eqs.~\eqref{lemma_qc_approx},~\eqref{eq_qi_rho1} and~\eqref{eq_qi_drho}, simple algebra shows that the quantum
  Chernoff bound for the state $\ket{\psi}_{\text{TMSV}}$ and a
  thermal environment with mean photon number $\nE$ has
  \begin{equation}
    \label{eq:kappaQ}
\kappa_\q =  1 -\frac{ r^2}{4 \nE}G \,,
\end{equation}
where 
\begin{subequations}
\begin{gather}
G = \sum_n (n+1) \mathcal{M}\left(c_n^2, \frac{\nE}{1+\nE} c_{n+1}^2\right) \,, \\
\mathcal{M}(x,y) = \frac{4 x y}{\left(\sqrt{ x}+\sqrt{y}\right)^2} \,.
\end{gather}
\end{subequations}
It can be easily verified that $\mathcal{M}(x,y)$ satisfies the
conditions for being a mean function, and is also concave.
Lemma \ref{lemma_epr_mean} then implies that the TMSV state maximizes
$G$, and hence minimizes the quantum Chernoff bound, which completes
the proof.\end{prooft2}

\noindent Substituting the coefficients $c_n$ for the TMSV
into~\eqref{eq:kappaQ}, the Chernoff bound for the TMSV in the limit of small $r$
has
\begin{equation}
  \kappa_\q=1-\frac{r^2 \nP \left( 1+\nE\right)}{\left[ 1+\nE +\sqrt{\nE\nP\frac{(1+\nE)}{1+\nP}}\right]^2}\,.
\end{equation}
Unlike the single-mode probe, for the two-mode probe, the optimal
state for individual measurement is not the same as the optimal probe
for collective measurement. In this case, we find that the TMSV state
does not minimize the Helstrom bound. Instead a numerical optimization
indicates that there exist a  different state that outperforms it.

\section{Quantum advantage comparisons} We can now compare the
performance of single-mode and two-mode QI schemes. In the limit of
small $r$, the quantum advantage ratio
\begin{equation}
  \frac{\log \kappa_\q}{\log \kappa_\cl} = \left[\frac{1+\nE+\sqrt{\nE(1+\nE)}}{1+\nE+\sqrt{\frac{\nP \nE(1+\nE)}{1+\nP}}} \right]^2\,,
\end{equation}
gives the multiple of additional copies required by single-mode QI to
achieve the same error probability scaling
as two-mode QI. This is plotted in Fig.~\ref{fig_qiadvantage}. The
figure demonstrates that the quantum advantage is greatest for low
probe energies and high environment noise. A maximum advantage of 4 is
achieved when $\nP \to 0$ and $ \nE \gg 1$~\cite{Tan_etal_PRL_08}. As $\nE$
  becomes smaller, the maximum quantum advantage, which occurs when
  $\nP\to 0$, decreases and until finally there is no quantum advantage when
  the environment is in the vacuum state~\cite{Shapiro_Lloyd_NJP_09}.
  When $\nP \gg 1$, $\kappa_\text{q}\to \kappa_c$ regardless
  of $\nE$, and the the coherent state probe performs almost as well
  as the TMSV probe.  In this regime, there is no advantage from using
  a two-mode probe even when the environment noise is large.

\begin{figure}[t]
\includegraphics[width=\columnwidth]{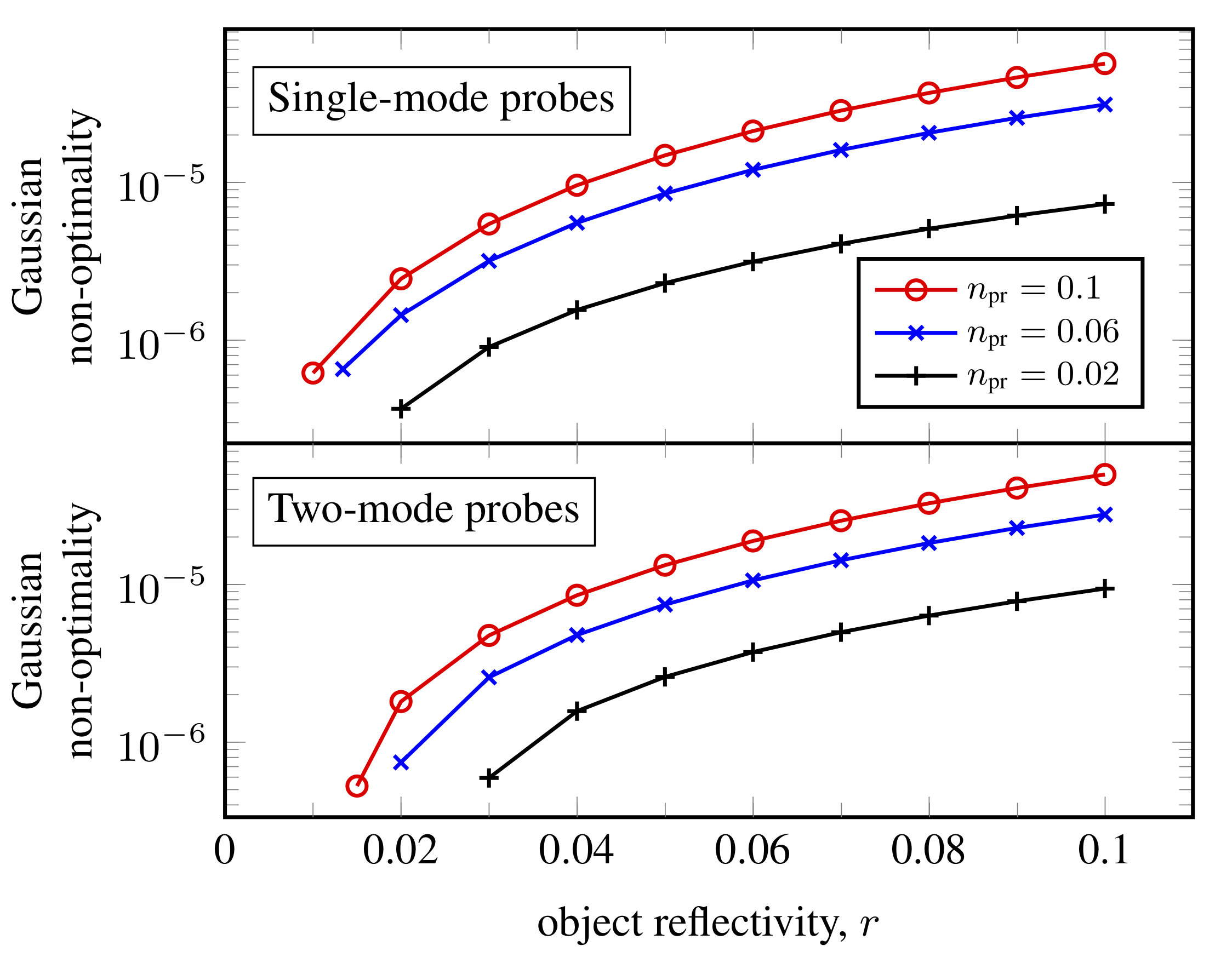}\caption{{\bf Comparison between the performance of the optimal
    (non-Gaussian) probe for QI and the Gaussian probe.}  The top
  figure shows the single mode Gaussian
    non-optimality for three different values of $\nP=0.1$,
  $0.06$ and $0.02$.  The bottom figure shows the same quantity
 for two mode states. The coherent and
  squeezed-vacuum probes are only optimal at $\ar=0$. However, they
  still remain close to optimal even when $\ar>0$.  The optimal probe
  states for $\ar>0$ are non-Gaussian and were computed
  numerically. The environment mean photon number is set to $\nE=0.5$
  in all the simulations.}
\label{fig_gaussperf}
\end{figure}

We have found the optimal states in the limit of zero object
reflectivity, but what about for non-zero $\ar$? It turns out that for
non-zero object reflectivity the coherent state and TMSV are no longer
the optimal probe states for single-mode and two-mode QI
respectively. This is not unexpected since at $\ar=1$ the problem
reduces to discrete-variable illumination with a constraint on the
probe energy. But how good are the Gaussian states? To answer this
question we compare the coherent state and TMSV state to an optimal
state found by numerical optimization~\footnote{Numerical optimization
  was carried out using standard techniques and programmes in
  MATLAB. The coherent and TMSV states were used as starting points
  for the density matrices in the single and two mode case
  respectively. These states were then subject to the operations
  representing a quantum illumination channel and the subsequent error
  probabilities were computed using Eq.~\eqref{kapeqn}. The error
  probability was then minimised over physical density matrices,
  subject to the energy constraint, using fmincon. For this reason,
  the solutions found might correspond to local minima instead of the
  global minima}. To this end we define the following quantity as the
\textit{Gaussian non-optimality}
$  1-\log \kappa_\text{gaus}/\log \kappa_\text{opt}$,
where $\kappa_\text{gaus}$ is the quantum chernoff exponent for
the optimal Gaussian states and $\kappa_\text{opt}$ is the quantum
chernoff exponent for the numerically optimimised state. The closer
this quantity is to zero the closer Gaussian states are to optimal.
This is plotted for both single and two mode states in
Fig.~\ref{fig_gaussperf}, for a particular choice of environment
energy, $\nE=0.5$, and a range of probe energies $\nP$ between $0.02$
and $0.1$. This shows that the performance of the Gaussian states is
close to optimal for small object reflectivities.  When the probe mean
photon number is small, the Gaussian state will always perform almost
as well as the optimal state. This is because as the mean photon
number approaches zero, all the states of the given energy look
similar, since they are mostly dominated by the zero and one photon
number components.

\section{Discussion and Conclusions}
Our findings complement existing results in the literature where
coherent states and TMSV probes were considered, but they were not
known or claimed to be optimal~\cite{Tan_etal_PRL_08,
  Guha_Erkmen_PRA_09, Lee_Ihn_Kim_arxiv_20}. For example, our results
are consistent with Ref.~\cite{Nair_Gu_arxiv_20} where a lower bound on the
probability of error was derived which does not rule out the TMSV
being optimal. These results also nicely
supplement Ref.~\cite{DePalma_Borregaard_PRA_18} which showed that the
coherent state and the TMSV are optimal in an asymmetric
version of continuous-variable QI where the goal is to minimize the
probability of a missed detection for a given probability of a false
alarm~\cite{Spedalieri_Braunstein_PRA_14, Wilde_etal_PRL_17}.

In this work we proved that in the limit of zero object
reflectivity, the coherent state is the
optimal probe state for single-mode QI whereas the two-mode
squeezed-vacuum state is the optimal two-mode probe. The low reflectivity limit is particularly relevant for microwave QI. We have also demonstrated that these states remain close to optimal for non-zero reflectivities. This result establishes the possibility of continuous-variable states being used as a viable, near-term, close to optimal platform for QI.
 Future research directions include finding the optimal probe states for
non-zero reflectivity and investigating the possible application of
multipartite entangled states to QI.

This research is supported by the Australian Research Council (ARC) under the Centre of Excellence
for Quantum Computation and Communication Technology (CE110001027), the Singapore Ministry of Education Tier 1 grant MOE2019-
T1-002-015, and National Research Foundation Fellowship NRF-NRFF2016-02. We acknowledge funding from the Defence Science and Technology Group.
\appendix

\section{Form of optimal probe two mode quantum illumination}
\label{th_diag}
We claim that there exists an optimal probe state for two mode quantum illumination of the form
\begin{equation}
\label{eq_optqiprobe}
\ket{\psi} = \sum_n c_n \ket{n, n} \,,
\end{equation}
where $c_n \geqslant 0$.

\begin{proofb}
The most general probe state is 
\begin{equation}
\label{general}
\ket{\Psi'}=\sum_n c_n\ket{n, \phi_n} ,
\end{equation}
where ${\phi_n}$ are a set of states that aren't necessarily orthogonal. $c_n$ can be taken to be real since any phase can be absorbed into $\ket{\phi_n}$. Let $\rho_0$ and $\rho_1$ be the possible detected states when the probe is (\ref{eq_optqiprobe}) and $\rho'_0$ and $\rho'_1$ be the possible detected states when the probe is (\ref{general}). The existence of a quantum operation which simultaneously transforms $\rho_0$ to $\rho'_0$ and $\rho_1$ to $\rho'_1$ would prove that a probe given by Eq. (\ref{eq_optqiprobe}) can perform no worse than a probe given by Eq. (\ref{general}) proving the theorem. This relies on the assumption that the figure of merit is one such that a quantum operation does not make two states more distinguishable. This is necessarily true of a figure that involves optimisation over measurements, such as probability of error, as any such operation can be included in the measurement. It is also true of the Holevo information, which can be understood because the Holevo information can be obtained by a measurement in the limit of infinite copies of the state. 

For a perfectly reflecting object with $r=1$,
\begin{equation}
\rho_1=\sum_{n,m}c_nc_m\ket{n, n}\bra{m, m}\;,
\end{equation}
\begin{equation}
\rho'_1=\sum_{n,m}c_nc_m\ket{n,\phi_n}\bra{m,\phi_m}\;,
\end{equation}
\begin{equation}
\rho_0=\sum_k \lambda_k\ket{k}\bra{k}\otimes\sum_m |c_m|^2\ket{m}\bra{m}\;,
\end{equation}
\begin{equation}
\rho'_0=\sum_k \lambda_k\ket{k}\bra{k}\otimes\sum_m |c_m|^2\ket{\phi_m}\bra{\phi_m} \;,
\end{equation}
where $\lambda_k$ depends on the number of thermal photons in the environment. We can then do the following quantum operation to transform from $\rho_0$ to $\rho'_0$ and $\rho_1$ to $\rho'_1$:
\begin{enumerate}
\item Perform the measurement with POVM elements $\Pi_0=I-\Pi_1$, $\Pi_1=\sum_{n}\ket{n, n}\bra{n, n}$.
\item If the measurement outcome is 1, do a unitary transformation $U$ such that $U\ket{n, n}=\ket{n, \phi_n}$ for all n.
\item If the measurement outcome is 0, do the non-unitary operation $O(\ket{n}_B)=\ket{\phi_n}_B$.
\end{enumerate} 
The measurement in step 1 checks whether there are the same number of photons in mode $A$ and mode $B$. Since this is always true for $\rho_1$, measuring $\rho_1$ will leave it unchanged and the result will be 1. Measuring $\rho_0$ will result in a state with $n=m$ if the measurement result is 1, and $n\neq m$ if the measurement result is 0.

The operation defined in step 2 is unitary since it transforms one set of orthogonal states into another set of orthogonal states. Note that this unitary is not uniquely defined, but this does not matter. Performing this unitary operation on $\rho_1$ will transform it to $\rho'_1$. This unitary also transforms the pure states that make up $\rho_0$ for which $n=m$ into the corresponding pure states of $\rho'_0$. 

The measurement outcome of 1 means that the state must have been $\rho_0$. The operation defined in step 3 corresponds to measuring the number of photons in mode $B$, and replacing the state with $\phi_n$ where $n$ is the number of photons measured. This operation will transform the remaining pure states of $\rho_0$ with $n\neq m$ into the corresponding $\rho'_0$ pure states.

For all other $r$ the beamsplitter operation makes the problem more complicated. Nevertheless we can still find the necessary transformation. The states are
\begin{equation}
\rho_1=\sum_{n,m,k}c_nc_m\lambda_k\text{tr}_{\text{env}}\{B(r)\ket{n, k}\bra{m, k}_{AE}B^{\dagger}(r)\}\ket{n}\bra{m}_B
\end{equation}
\begin{equation}
\rho'_1=\sum_{n,m,k}c_nc_m\lambda_k\text{tr}_{\text{env}}\{B(r)\ket{n, k}\bra{m, k}_{AE}B^{\dagger}(r)\}\ket{\phi_n}\bra{\phi_m}_B
\end{equation}
\begin{equation}
\rho_0=\sum_k \lambda_k\ket{k}\bra{k}\otimes\sum_m |c_m|^2\ket{m}\bra{m}
\end{equation}
\begin{equation}
\rho'_0=\sum_k \lambda_k\ket{k}\bra{k}\otimes\sum_m |c_m|^2\ket{\phi_m}\bra{\phi_m} .
\end{equation}
The pure states that make up $\rho_1$ for $\nE=0$ where $\nE$ is the number of photons in the environment before the beam splitter will be
\begin{equation}
\label{m01}
\ket{0,0}_{AB}\ket{0}_E+\ket{1,1}_{AB}\ket{0}_E+\ket{2,2}_{AB}\ket{0}_E+....
\end{equation}
\begin{equation}
\label{m02}
\ket{0,1}_{AB}\ket{1}_E+\ket{1,2}_{AB}\ket{1}_E+\ket{2,3}_{AB}\ket{1}_E+....
\end{equation}
\begin{equation}
\ket{0,2}_{AB}\ket{2}_E+\ket{1,3}_{AB}\ket{2}_E+\ket{2,4}_{AB}\ket{2}_E+....
\end{equation}
and so on. For $\nE=1$ we have 
\begin{equation}
\label{m11}
\ket{1,0}_{AB}\ket{0}_E+\ket{2,1}_{AB}\ket{0}_E+\ket{3,2}_{AB}\ket{0}_E+....
\end{equation}
\begin{equation}
\label{m12}
\ket{0,0}_{AB}\ket{1}_E+\ket{1,1}_{AB}\ket{1}_E+\ket{2,2}_{AB}\ket{1}_E+....
\end{equation}
\begin{equation}
\label{m13}
\ket{0,1}_{AB}\ket{2}_E+\ket{1,2}_{AB}\ket{2}_E+\ket{2,3}_{AB}\ket{2}_E+....
\end{equation}
and so on, where the environment component is traced out and each term in the expressions are multiplied by some number which we have neglected because it is irrelevant for the discussion. The pure states that make up $\rho'_1$ are the same except $\ket{n}_B$ is replaced with $\ket{\phi_B}$. The important part to note about $\rho_1$ is that we can measure the difference between the number of photons in mode $A$ and mode $B$ without disturbing any of the pure states. For example states (\ref{m01}) and (\ref{m12}) have the same number of photons in modes $A$ and $B$, mode $A$ of states (\ref{m02}) and (\ref{m13}) lost one photon to the environment and for state (\ref{m11}) mode $A$ gained one photon from the environment. The POVM elements that make up this measurement are 
\begin{equation}
\Pi_0=\sum_n\ket{n,n}\bra{n,n} \;,
\end{equation}
\begin{equation}
\Pi_k=\sum_n\ket{n+k,n}\bra{n+k,n}\text{   for }k\geq 1\;,
\end{equation}
\begin{equation}
\Pi_{-k}=\sum_n\ket{n,n+k}\bra{n,n+k}\text{   for } k\geq 1\;,
\end{equation}
which form a valid measurement since $\sum_{j=-\infty}^{\infty}\Pi_j=I$. If this measurement is done on $\rho_1$, we will be left with a mixture of the pure states that have the photon difference corresponding to the measurement outcome. Performing a corresponding unitary operation defined by
\begin{equation}
U_0\ket{n,n}=\ket{n,\phi_n}\;,
\end{equation}
\begin{equation}
U_k\ket{n+k,n}=\ket{n+k,\phi_n} \text{    for } k\geq1\;,
\end{equation}
\begin{equation}
U_{-k}\ket{n,n+k}=\ket{n,\phi_{n+k}} \text{    for } k\geq1\;,
\end{equation}
converts the pure states of $\rho_1$ into the pure states $\rho'_1$. The quantum operation defined by the measurement and unitary transformations also converts $\rho_0$ to $\rho'_0$. The existence of this operation proves our claim.
\end{proofb}

\section{Proof of lemma 1}
\begin{lemma}
\label{lemma_epr_mean}
The EPR state $\ket{\psi} = \sum_n c_n \ket{n}\ket{n}$ with $c_n = \sqrt{\bar{n}^n/(\bar{n}+1)^{n+1}}$, is a local optima for the maximisation of
\begin{equation}
\label{eq_meanmax}
G=\sum_n (n+1) \mathcal{M}(c_n^2, g c_{n+1}^2) \,,
\end{equation} 
where $\mathcal{M}(x,y)$ is a mean function and $g$ is a positive constant. If $\mathcal{M}(x,y)$ is concave it is globally optimal.
\end{lemma}

\begin{proofb}
A mean function $m(x,y)$ is a mean function if it can be written as $m(x,y)=yf(\frac{x}{y})$ where $f$ has the following properties:
\begin{enumerate}
\item $f$ is monotone increasing, i.e. $x\geq y$ implies $f(x)\geq f(y)$.
\item $xf(x^{-1})=f(x)$. This ensures $m(x,y)$ is symmetric in $x$ and $y$.
\item $f(1)=1$.
\end{enumerate}
We now use property 2 of a mean function and take the derivative 
\begin{equation}
f'(x)=f(x^{-1})+xf'(x^{-1}).
\end{equation}
Since 
\begin{equation}
G=\sum_{n=0}^{\infty} (n+1) c_{n+1}^2 f\left(\frac{c_n^2}{c_{n+1}^2}\right)=\sum_{n=1}^\infty nc_{n-1}f\left(\frac{c_n^2}{c_{n-1}^2}\right)\;.
\end{equation}
The derivative is
\begin{multline}
\frac{\partial G}{\partial c_k}=(k+1)2c_k\left(f\left(\frac{c_{k+1}^2}{c_k^2}\right)+\frac{c_k^2}{c_{k+1}^2}f'\left(\frac{c_{k+1}^2}{c_{k}^2}\right)\right) \\ +2kc_k\left(f\left(\frac{c_{k-1}^2}{c_k^2}\right)+\frac{c_k^2}{c_{k-1}^2}f'\left(\frac{c_{k-1}^2}{c_k^2}\right)\right) .
\end{multline}
Using the normalisation and energy constraints as we did in the classical case we get the following condition for the optimal point
\begin{equation}
\frac{\partial G}{\partial c_k}+2c_k\mu_1+2c_kk\mu_2=0 ,
\end{equation}
where $\mu_1$ and $\mu_2$ are Lagrange multipliers as before. This equation can be satisfied if the term multiplied by $2c_k$ equals zero and if the term multiplied by $2c_kk$ equals zero, i.e.
\begin{equation}
f\left(\frac{c_{k-1}^2}{c_k^2}\right)+\frac{c_k^2}{c_{k-1}^2}f'\left(\frac{c_{k-1}^2}{c_k^2}\right)+\mu_1=0 
\end{equation}
\begin{equation}
f\left(\frac{c_{k+1}^2}{c_k^2}\right)+\frac{c_k^2}{c_{k+1}^2}f'\left(\frac{c_{k+1}^2}{c_k^2}\right)+f\left(\frac{c_{k-1}^2}{c_k^2}\right)+\frac{c_k^2}{c_{k-1}^2}f'\left(\frac{c_{k-1}^2}{c_k^2}\right)+\mu_2=0 .
\end{equation}
By choosing $c_k= \sqrt{\bar{n}^k/(\bar{n}+1)^{k+1}}$, the dependency of the above equations on $k$ is removed, hence it is possible to find $\mu_1$ and $\mu_2$ which satisfy the equations. Thus the two mode squeezed vacuum is optimal. 
\end{proofb}

\bibliography{bibliography}

\begin{thebibliography}{36}%
\makeatletter
\providecommand \@ifxundefined [1]{%
 \@ifx{#1\undefined}
}%
\providecommand \@ifnum [1]{%
 \ifnum #1\expandafter \@firstoftwo
 \else \expandafter \@secondoftwo
 \fi
}%
\providecommand \@ifx [1]{%
 \ifx #1\expandafter \@firstoftwo
 \else \expandafter \@secondoftwo
 \fi
}%
\providecommand \natexlab [1]{#1}%
\providecommand \enquote  [1]{``#1''}%
\providecommand \bibnamefont  [1]{#1}%
\providecommand \bibfnamefont [1]{#1}%
\providecommand \citenamefont [1]{#1}%
\providecommand \href@noop [0]{\@secondoftwo}%
\providecommand \href [0]{\begingroup \@sanitize@url \@href}%
\providecommand \@href[1]{\@@startlink{#1}\@@href}%
\providecommand \@@href[1]{\endgroup#1\@@endlink}%
\providecommand \@sanitize@url [0]{\catcode `\\12\catcode `\$12\catcode
  `\&12\catcode `\#12\catcode `\^12\catcode `\_12\catcode `\%12\relax}%
\providecommand \@@startlink[1]{}%
\providecommand \@@endlink[0]{}%
\providecommand \url  [0]{\begingroup\@sanitize@url \@url }%
\providecommand \@url [1]{\endgroup\@href {#1}{\urlprefix }}%
\providecommand \urlprefix  [0]{URL }%
\providecommand \Eprint [0]{\href }%
\providecommand \doibase [0]{http://dx.doi.org/}%
\providecommand \selectlanguage [0]{\@gobble}%
\providecommand \bibinfo  [0]{\@secondoftwo}%
\providecommand \bibfield  [0]{\@secondoftwo}%
\providecommand \translation [1]{[#1]}%
\providecommand \BibitemOpen [0]{}%
\providecommand \bibitemStop [0]{}%
\providecommand \bibitemNoStop [0]{.\EOS\space}%
\providecommand \EOS [0]{\spacefactor3000\relax}%
\providecommand \BibitemShut  [1]{\csname bibitem#1\endcsname}%
\let\auto@bib@innerbib\@empty
\bibitem [{\citenamefont {Lloyd}(2008)}]{Lloyd_S_08}%
  \BibitemOpen
  \bibfield  {author} {\bibinfo {author} {\bibfnamefont {S.}~\bibnamefont
  {Lloyd}},\ }\href {\doibase 10.1126/science.1160627} {\bibfield  {journal}
  {\bibinfo  {journal} {Science}\ }\textbf {\bibinfo {volume} {321}},\ \bibinfo
  {pages} {1463} (\bibinfo {year} {2008})}\BibitemShut {NoStop}%
\bibitem [{\citenamefont {Tan}\ \emph {et~al.}(2008)\citenamefont {Tan},
  \citenamefont {Erkmen}, \citenamefont {Giovannetti}, \citenamefont {Guha},
  \citenamefont {Lloyd}, \citenamefont {Maccone}, \citenamefont {Pirandola},\
  and\ \citenamefont {Shapiro}}]{Tan_etal_PRL_08}%
  \BibitemOpen
  \bibfield  {author} {\bibinfo {author} {\bibfnamefont {S.-H.}\ \bibnamefont
  {Tan}}, \bibinfo {author} {\bibfnamefont {B.~I.}\ \bibnamefont {Erkmen}},
  \bibinfo {author} {\bibfnamefont {V.}~\bibnamefont {Giovannetti}}, \bibinfo
  {author} {\bibfnamefont {S.}~\bibnamefont {Guha}}, \bibinfo {author}
  {\bibfnamefont {S.}~\bibnamefont {Lloyd}}, \bibinfo {author} {\bibfnamefont
  {L.}~\bibnamefont {Maccone}}, \bibinfo {author} {\bibfnamefont
  {S.}~\bibnamefont {Pirandola}}, \ and\ \bibinfo {author} {\bibfnamefont
  {J.~H.}\ \bibnamefont {Shapiro}},\ }\href
  {https://doi.org/10.1103/PhysRevLett.101.253601} {\bibfield  {journal}
  {\bibinfo  {journal} {Phys. Rev. Lett.}\ }\textbf {\bibinfo {volume} {101}},\
  \bibinfo {pages} {253601} (\bibinfo {year} {2008})}\BibitemShut {NoStop}%
\bibitem [{\citenamefont {Shapiro}\ and\ \citenamefont
  {Lloyd}(2009)}]{Shapiro_Lloyd_NJP_09}%
  \BibitemOpen
  \bibfield  {author} {\bibinfo {author} {\bibfnamefont {J.~H.}\ \bibnamefont
  {Shapiro}}\ and\ \bibinfo {author} {\bibfnamefont {S.}~\bibnamefont
  {Lloyd}},\ }\href {\doibase 10.1088/1367-2630/11/6/063045} {\bibfield
  {journal} {\bibinfo  {journal} {New J. Phys.}\ }\textbf {\bibinfo {volume}
  {11}},\ \bibinfo {pages} {063045} (\bibinfo {year} {2009})}\BibitemShut
  {NoStop}%
\bibitem [{\citenamefont {Zhuang}\ \emph {et~al.}(2017)\citenamefont {Zhuang},
  \citenamefont {Zhang},\ and\ \citenamefont
  {Shapiro}}]{Zhuang_Zhang_Shapiro_PRL_17}%
  \BibitemOpen
  \bibfield  {author} {\bibinfo {author} {\bibfnamefont {Q.}~\bibnamefont
  {Zhuang}}, \bibinfo {author} {\bibfnamefont {Z.}~\bibnamefont {Zhang}}, \
  and\ \bibinfo {author} {\bibfnamefont {J.~H.}\ \bibnamefont {Shapiro}},\
  }\href {\doibase 10.1103/PhysRevLett.118.040801} {\bibfield  {journal}
  {\bibinfo  {journal} {Phys. Rev. Lett.}\ }\textbf {\bibinfo {volume} {118}},\
  \bibinfo {pages} {040801} (\bibinfo {year} {2017})}\BibitemShut {NoStop}%
\bibitem [{\citenamefont {Zhang}\ \emph {et~al.}(2014)\citenamefont {Zhang},
  \citenamefont {Zou}, \citenamefont {Shi}, \citenamefont {Guo},\ and\
  \citenamefont {Guo}}]{Zhang_etal_PRA_14}%
  \BibitemOpen
  \bibfield  {author} {\bibinfo {author} {\bibfnamefont {S.}~\bibnamefont
  {Zhang}}, \bibinfo {author} {\bibfnamefont {X.}~\bibnamefont {Zou}}, \bibinfo
  {author} {\bibfnamefont {J.}~\bibnamefont {Shi}}, \bibinfo {author}
  {\bibfnamefont {J.}~\bibnamefont {Guo}}, \ and\ \bibinfo {author}
  {\bibfnamefont {G.}~\bibnamefont {Guo}},\ }\href {\doibase
  10.1103/PhysRevA.90.052308} {\bibfield  {journal} {\bibinfo  {journal} {Phys.
  Rev. A}\ }\textbf {\bibinfo {volume} {90}},\ \bibinfo {pages} {052308}
  (\bibinfo {year} {2014})}\BibitemShut {NoStop}%
\bibitem [{\citenamefont {Nair}\ and\ \citenamefont
  {Gu}(2020)}]{Nair_Gu_arxiv_20}%
  \BibitemOpen
  \bibfield  {author} {\bibinfo {author} {\bibfnamefont {R.}~\bibnamefont
  {Nair}}\ and\ \bibinfo {author} {\bibfnamefont {M.}~\bibnamefont {Gu}},\
  }\href {https://arxiv.org/abs/2002.12252} {\bibfield  {journal} {\bibinfo
  {journal} {Preprint at arXiv:2002.12252}\ } (\bibinfo {year}
  {2020})}\BibitemShut {NoStop}%
\bibitem [{\citenamefont {Las~Heras}\ \emph {et~al.}(2017)\citenamefont
  {Las~Heras}, \citenamefont {Di~Candia}, \citenamefont {Fedorov},
  \citenamefont {Deppe}, \citenamefont {Sanz},\ and\ \citenamefont
  {Solano}}]{LasHeras_SR_17}%
  \BibitemOpen
  \bibfield  {author} {\bibinfo {author} {\bibfnamefont {U.}~\bibnamefont
  {Las~Heras}}, \bibinfo {author} {\bibfnamefont {R.}~\bibnamefont
  {Di~Candia}}, \bibinfo {author} {\bibfnamefont {K.~G.}\ \bibnamefont
  {Fedorov}}, \bibinfo {author} {\bibfnamefont {F.}~\bibnamefont {Deppe}},
  \bibinfo {author} {\bibfnamefont {M.}~\bibnamefont {Sanz}}, \ and\ \bibinfo
  {author} {\bibfnamefont {E.}~\bibnamefont {Solano}},\ }\href {\doibase
  10.1038/s41598-017-08505-w} {\bibfield  {journal} {\bibinfo  {journal} {Sci.
  Rep.}\ }\textbf {\bibinfo {volume} {7}},\ \bibinfo {pages} {9333} (\bibinfo
  {year} {2017})}\BibitemShut {NoStop}%
\bibitem [{\citenamefont {Shapiro}(2009)}]{Shapiro_PRA_09}%
  \BibitemOpen
  \bibfield  {author} {\bibinfo {author} {\bibfnamefont {J.~H.}\ \bibnamefont
  {Shapiro}},\ }\href {\doibase 10.1103/PhysRevA.80.022320} {\bibfield
  {journal} {\bibinfo  {journal} {Phys. Rev. A}\ }\textbf {\bibinfo {volume}
  {80}},\ \bibinfo {pages} {022320} (\bibinfo {year} {2009})}\BibitemShut
  {NoStop}%
\bibitem [{\citenamefont {Zhang}\ \emph {et~al.}(2015)\citenamefont {Zhang},
  \citenamefont {Mouradian}, \citenamefont {Wong},\ and\ \citenamefont
  {Shapiro}}]{Zhang_etal_PRL_15}%
  \BibitemOpen
  \bibfield  {author} {\bibinfo {author} {\bibfnamefont {Z.}~\bibnamefont
  {Zhang}}, \bibinfo {author} {\bibfnamefont {S.}~\bibnamefont {Mouradian}},
  \bibinfo {author} {\bibfnamefont {F.~N.~C.}\ \bibnamefont {Wong}}, \ and\
  \bibinfo {author} {\bibfnamefont {J.~H.}\ \bibnamefont {Shapiro}},\ }\href
  {\doibase 10.1103/PhysRevLett.114.110506} {\bibfield  {journal} {\bibinfo
  {journal} {Phys. Rev. Lett.}\ }\textbf {\bibinfo {volume} {114}},\ \bibinfo
  {pages} {110506} (\bibinfo {year} {2015})}\BibitemShut {NoStop}%
\bibitem [{\citenamefont {Lopaeva}\ \emph {et~al.}(2013)\citenamefont
  {Lopaeva}, \citenamefont {{Ruo Berchera}}, \citenamefont {Degiovanni},
  \citenamefont {Olivares}, \citenamefont {Brida},\ and\ \citenamefont
  {Genovese}}]{Lopaeva_etal_PRL_13}%
  \BibitemOpen
  \bibfield  {author} {\bibinfo {author} {\bibfnamefont {E.~D.}\ \bibnamefont
  {Lopaeva}}, \bibinfo {author} {\bibfnamefont {I.}~\bibnamefont {{Ruo
  Berchera}}}, \bibinfo {author} {\bibfnamefont {I.~P.}\ \bibnamefont
  {Degiovanni}}, \bibinfo {author} {\bibfnamefont {S.}~\bibnamefont
  {Olivares}}, \bibinfo {author} {\bibfnamefont {G.}~\bibnamefont {Brida}}, \
  and\ \bibinfo {author} {\bibfnamefont {M.}~\bibnamefont {Genovese}},\ }\href
  {\doibase 10.1103/PhysRevLett.110.153603} {\bibfield  {journal} {\bibinfo
  {journal} {Phys. Rev. Lett.}\ }\textbf {\bibinfo {volume} {110}},\ \bibinfo
  {pages} {153603} (\bibinfo {year} {2013})}\BibitemShut {NoStop}%
\bibitem [{\citenamefont {England}\ \emph {et~al.}(2019)\citenamefont
  {England}, \citenamefont {Balaji},\ and\ \citenamefont
  {Sussman}}]{England_Balaji_Sussman_PRA_19}%
  \BibitemOpen
  \bibfield  {author} {\bibinfo {author} {\bibfnamefont {D.~G.}\ \bibnamefont
  {England}}, \bibinfo {author} {\bibfnamefont {B.}~\bibnamefont {Balaji}}, \
  and\ \bibinfo {author} {\bibfnamefont {B.~J.}\ \bibnamefont {Sussman}},\
  }\href {\doibase 10.1103/PhysRevA.99.023828} {\bibfield  {journal} {\bibinfo
  {journal} {Phys. Rev. A}\ }\textbf {\bibinfo {volume} {99}},\ \bibinfo
  {pages} {023828} (\bibinfo {year} {2019})}\BibitemShut {NoStop}%
\bibitem [{\citenamefont {Aguilar}\ \emph {et~al.}(2019)\citenamefont
  {Aguilar}, \citenamefont {de~Souza}, \citenamefont {Gomes}, \citenamefont
  {Thompson}, \citenamefont {Gu}, \citenamefont {C\'eleri},\ and\ \citenamefont
  {Walborn}}]{Aguilar_etal_PRA_19}%
  \BibitemOpen
  \bibfield  {author} {\bibinfo {author} {\bibfnamefont {G.~H.}\ \bibnamefont
  {Aguilar}}, \bibinfo {author} {\bibfnamefont {M.~A.}\ \bibnamefont
  {de~Souza}}, \bibinfo {author} {\bibfnamefont {R.~M.}\ \bibnamefont {Gomes}},
  \bibinfo {author} {\bibfnamefont {J.}~\bibnamefont {Thompson}}, \bibinfo
  {author} {\bibfnamefont {M.}~\bibnamefont {Gu}}, \bibinfo {author}
  {\bibfnamefont {L.~C.}\ \bibnamefont {C\'eleri}}, \ and\ \bibinfo {author}
  {\bibfnamefont {S.~P.}\ \bibnamefont {Walborn}},\ }\href {\doibase
  10.1103/PhysRevA.99.053813} {\bibfield  {journal} {\bibinfo  {journal} {Phys.
  Rev. A}\ }\textbf {\bibinfo {volume} {99}},\ \bibinfo {pages} {053813}
  (\bibinfo {year} {2019})}\BibitemShut {NoStop}%
\bibitem [{\citenamefont {Zhang}\ \emph {et~al.}(2013)\citenamefont {Zhang},
  \citenamefont {Tengner}, \citenamefont {Zhong}, \citenamefont {Wong},\ and\
  \citenamefont {Shapiro}}]{Zhang_etal_PRL_13}%
  \BibitemOpen
  \bibfield  {author} {\bibinfo {author} {\bibfnamefont {Z.}~\bibnamefont
  {Zhang}}, \bibinfo {author} {\bibfnamefont {M.}~\bibnamefont {Tengner}},
  \bibinfo {author} {\bibfnamefont {T.}~\bibnamefont {Zhong}}, \bibinfo
  {author} {\bibfnamefont {F.~N.~C.}\ \bibnamefont {Wong}}, \ and\ \bibinfo
  {author} {\bibfnamefont {J.~H.}\ \bibnamefont {Shapiro}},\ }\href
  {http://link.aps.org/doi/10.1103/PhysRevLett.111.010501} {\bibfield
  {journal} {\bibinfo  {journal} {Phys. Rev. Lett.}\ }\textbf {\bibinfo
  {volume} {111}},\ \bibinfo {pages} {010501} (\bibinfo {year}
  {2013})}\BibitemShut {NoStop}%
\bibitem [{\citenamefont {Barzanjeh}\ \emph {et~al.}(2015)\citenamefont
  {Barzanjeh}, \citenamefont {Guha}, \citenamefont {Weedbrook}, \citenamefont
  {Vitali}, \citenamefont {Shapiro},\ and\ \citenamefont
  {Pirandola}}]{Barzanjeh_etal_PRL_15}%
  \BibitemOpen
  \bibfield  {author} {\bibinfo {author} {\bibfnamefont {S.}~\bibnamefont
  {Barzanjeh}}, \bibinfo {author} {\bibfnamefont {S.}~\bibnamefont {Guha}},
  \bibinfo {author} {\bibfnamefont {C.}~\bibnamefont {Weedbrook}}, \bibinfo
  {author} {\bibfnamefont {D.}~\bibnamefont {Vitali}}, \bibinfo {author}
  {\bibfnamefont {J.~H.}\ \bibnamefont {Shapiro}}, \ and\ \bibinfo {author}
  {\bibfnamefont {S.}~\bibnamefont {Pirandola}},\ }\href
  {http://link.aps.org/doi/10.1103/PhysRevLett.114.080503} {\bibfield
  {journal} {\bibinfo  {journal} {Phys. Rev. Lett.}\ }\textbf {\bibinfo
  {volume} {114}},\ \bibinfo {pages} {080503} (\bibinfo {year}
  {2015})}\BibitemShut {NoStop}%
\bibitem [{\citenamefont {Xiong}\ \emph {et~al.}(2017)\citenamefont {Xiong},
  \citenamefont {Li}, \citenamefont {Wang},\ and\ \citenamefont
  {Zhou}}]{Xiong_etal_AP_17}%
  \BibitemOpen
  \bibfield  {author} {\bibinfo {author} {\bibfnamefont {B.}~\bibnamefont
  {Xiong}}, \bibinfo {author} {\bibfnamefont {X.}~\bibnamefont {Li}}, \bibinfo
  {author} {\bibfnamefont {X.-Y.}\ \bibnamefont {Wang}}, \ and\ \bibinfo
  {author} {\bibfnamefont {L.}~\bibnamefont {Zhou}},\ }\href {\doibase
  https://doi.org/10.1016/j.aop.2017.08.024} {\bibfield  {journal} {\bibinfo
  {journal} {Ann. Phys.}\ }\textbf {\bibinfo {volume} {385}},\ \bibinfo {pages}
  {757 } (\bibinfo {year} {2017})}\BibitemShut {NoStop}%
\bibitem [{\citenamefont {Barzanjeh}\ \emph {et~al.}(2020)\citenamefont
  {Barzanjeh}, \citenamefont {Pirandola}, \citenamefont {Vitali},\ and\
  \citenamefont {Fink}}]{Barzanjeh_etal_SA_20}%
  \BibitemOpen
  \bibfield  {author} {\bibinfo {author} {\bibfnamefont {S.}~\bibnamefont
  {Barzanjeh}}, \bibinfo {author} {\bibfnamefont {S.}~\bibnamefont
  {Pirandola}}, \bibinfo {author} {\bibfnamefont {D.}~\bibnamefont {Vitali}}, \
  and\ \bibinfo {author} {\bibfnamefont {J.~M.}\ \bibnamefont {Fink}},\ }\href
  {\doibase 10.1126/sciadv.abb0451} {\bibfield  {journal} {\bibinfo  {journal}
  {Sci. Adv.}\ }\textbf {\bibinfo {volume} {6}} (\bibinfo {year} {2020}),\
  10.1126/sciadv.abb0451}\BibitemShut {NoStop}%
\bibitem [{\citenamefont {Luong}\ \emph {et~al.}(2019)\citenamefont {Luong},
  \citenamefont {Chang}, \citenamefont {Vadiraj}, \citenamefont {Damini},
  \citenamefont {Wilson},\ and\ \citenamefont {Balaji}}]{Luong_etal_IEEE_19}%
  \BibitemOpen
  \bibfield  {author} {\bibinfo {author} {\bibfnamefont {D.}~\bibnamefont
  {Luong}}, \bibinfo {author} {\bibfnamefont {C.~W.~S.}\ \bibnamefont {Chang}},
  \bibinfo {author} {\bibfnamefont {A.~M.}\ \bibnamefont {Vadiraj}}, \bibinfo
  {author} {\bibfnamefont {A.}~\bibnamefont {Damini}}, \bibinfo {author}
  {\bibfnamefont {C.~M.}\ \bibnamefont {Wilson}}, \ and\ \bibinfo {author}
  {\bibfnamefont {B.}~\bibnamefont {Balaji}},\ }\href@noop {} {\bibfield
  {journal} {\bibinfo  {journal} {IEEE Transactions on Aerospace and Electronic
  Systems}\ ,\ \bibinfo {pages} {1}} (\bibinfo {year} {2019})}\BibitemShut
  {NoStop}%
\bibitem [{\citenamefont {Luong}\ \emph {et~al.}(2018)\citenamefont {Luong},
  \citenamefont {Balaji}, \citenamefont {{Sandbo Chang}}, \citenamefont
  {{Ananthapadmanabha Rao}},\ and\ \citenamefont
  {Wilson}}]{Luong_etal_IEEE_18}%
  \BibitemOpen
  \bibfield  {author} {\bibinfo {author} {\bibfnamefont {D.}~\bibnamefont
  {Luong}}, \bibinfo {author} {\bibfnamefont {B.}~\bibnamefont {Balaji}},
  \bibinfo {author} {\bibfnamefont {C.~W.}\ \bibnamefont {{Sandbo Chang}}},
  \bibinfo {author} {\bibfnamefont {V.~M.}\ \bibnamefont {{Ananthapadmanabha
  Rao}}}, \ and\ \bibinfo {author} {\bibfnamefont {C.}~\bibnamefont {Wilson}},\
  }in\ \href@noop {} {\emph {\bibinfo {booktitle} {2018 International Carnahan
  Conference on Security Technology (ICCST)}}}\ (\bibinfo {year} {2018})\ pp.\
  \bibinfo {pages} {1--5}\BibitemShut {NoStop}%
\bibitem [{\citenamefont {Yung}\ \emph {et~al.}(2018)\citenamefont {Yung},
  \citenamefont {Meng},\ and\ \citenamefont {Zhao}}]{Yung_Meng_Zhao_arxiv_18}%
  \BibitemOpen
  \bibfield  {author} {\bibinfo {author} {\bibfnamefont {M.-H.}\ \bibnamefont
  {Yung}}, \bibinfo {author} {\bibfnamefont {F.}~\bibnamefont {Meng}}, \ and\
  \bibinfo {author} {\bibfnamefont {M.-J.}\ \bibnamefont {Zhao}},\ }\href
  {https://arxiv.org/abs/1801.07591} {\bibfield  {journal} {\bibinfo  {journal}
  {Preprint at arXiv:1801.07591}\ } (\bibinfo {year} {2018})}\BibitemShut
  {NoStop}%
\bibitem [{\citenamefont {Helstrom}(1969)}]{Helstrom_JSP_69}%
  \BibitemOpen
  \bibfield  {author} {\bibinfo {author} {\bibfnamefont {C.~W.}\ \bibnamefont
  {Helstrom}},\ }\href {\doibase 10.1007/BF01007479} {\bibfield  {journal}
  {\bibinfo  {journal} {J. Stat. Phys.}\ }\textbf {\bibinfo {volume} {1}},\
  \bibinfo {pages} {231} (\bibinfo {year} {1969})}\BibitemShut {NoStop}%
\bibitem [{\citenamefont {Audenaert}\ \emph {et~al.}(2007)\citenamefont
  {Audenaert}, \citenamefont {Calsamiglia}, \citenamefont {Mu\~noz Tapia},
  \citenamefont {Bagan}, \citenamefont {Masanes}, \citenamefont {Acin},\ and\
  \citenamefont {Verstraete}}]{Audenaert_etal_PRL_07}%
  \BibitemOpen
  \bibfield  {author} {\bibinfo {author} {\bibfnamefont {K.~M.~R.}\
  \bibnamefont {Audenaert}}, \bibinfo {author} {\bibfnamefont {J.}~\bibnamefont
  {Calsamiglia}}, \bibinfo {author} {\bibfnamefont {R.}~\bibnamefont {Mu\~noz
  Tapia}}, \bibinfo {author} {\bibfnamefont {E.}~\bibnamefont {Bagan}},
  \bibinfo {author} {\bibfnamefont {L.}~\bibnamefont {Masanes}}, \bibinfo
  {author} {\bibfnamefont {A.}~\bibnamefont {Acin}}, \ and\ \bibinfo {author}
  {\bibfnamefont {F.}~\bibnamefont {Verstraete}},\ }\href
  {https://link.aps.org/doi/10.1103/PhysRevLett.98.160501} {\bibfield
  {journal} {\bibinfo  {journal} {Phys. Rev. Lett.}\ }\textbf {\bibinfo
  {volume} {98}},\ \bibinfo {pages} {160501} (\bibinfo {year}
  {2007})}\BibitemShut {NoStop}%
\bibitem [{\citenamefont {Nussbaum}\ and\ \citenamefont
  {Szkoła}(2009)}]{Nussbaum_Szkola_AS_09}%
  \BibitemOpen
  \bibfield  {author} {\bibinfo {author} {\bibfnamefont {M.}~\bibnamefont
  {Nussbaum}}\ and\ \bibinfo {author} {\bibfnamefont {A.}~\bibnamefont
  {Szkoła}},\ }\href {\doibase 10.1214/08-AOS593} {\bibfield  {journal}
  {\bibinfo  {journal} {Ann. Statist.}\ }\textbf {\bibinfo {volume} {37}},\
  \bibinfo {pages} {1040} (\bibinfo {year} {2009})}\BibitemShut {NoStop}%
\bibitem [{\citenamefont {Bauer}(1958)}]{Bauer_ADM_58}%
  \BibitemOpen
  \bibfield  {author} {\bibinfo {author} {\bibfnamefont {H.}~\bibnamefont
  {Bauer}},\ }\href {\doibase 10.1007/BF01898615} {\bibfield  {journal}
  {\bibinfo  {journal} {Archiv der Mathematik}\ }\textbf {\bibinfo {volume}
  {9}},\ \bibinfo {pages} {389} (\bibinfo {year} {1958})}\BibitemShut {NoStop}%
\bibitem [{\citenamefont {Pardalos}\ and\ \citenamefont
  {Vavasis}(1991)}]{Pardalos_Vavasis_JGO_91}%
  \BibitemOpen
  \bibfield  {author} {\bibinfo {author} {\bibfnamefont {P.~M.}\ \bibnamefont
  {Pardalos}}\ and\ \bibinfo {author} {\bibfnamefont {S.~A.}\ \bibnamefont
  {Vavasis}},\ }\href {\doibase 10.1007/BF00120662} {\bibfield  {journal}
  {\bibinfo  {journal} {J. of Global Optim.}\ }\textbf {\bibinfo {volume}
  {1}},\ \bibinfo {pages} {15} (\bibinfo {year} {1991})}\BibitemShut {NoStop}%
\bibitem [{\citenamefont {Sahni}(1974)}]{Sahni_JC_74}%
  \BibitemOpen
  \bibfield  {author} {\bibinfo {author} {\bibfnamefont {S.}~\bibnamefont
  {Sahni}},\ }\href {https://doi.org/10.1137/0203021} {\bibfield  {journal}
  {\bibinfo  {journal} {SIAM J. Comp.}\ }\textbf {\bibinfo {volume} {3}},\
  \bibinfo {pages} {262} (\bibinfo {year} {1974})}\BibitemShut {NoStop}%
\bibitem [{sup()}]{supplmat}%
  \BibitemOpen
  \href@noop {} {}\bibinfo {note} {\text{See Supplemental Material} at link
  provided for details on the form of the optimal two mode probe (theorem 1)
  and the proof of lemma 1}\BibitemShut {NoStop}%
\bibitem [{\citenamefont {Sharma}\ \emph {et~al.}(2018)\citenamefont {Sharma},
  \citenamefont {Wilde}, \citenamefont {Adhikari},\ and\ \citenamefont
  {Takeoka}}]{Sharma_etal_NJP_18}%
  \BibitemOpen
  \bibfield  {author} {\bibinfo {author} {\bibfnamefont {K.}~\bibnamefont
  {Sharma}}, \bibinfo {author} {\bibfnamefont {M.~M.}\ \bibnamefont {Wilde}},
  \bibinfo {author} {\bibfnamefont {S.}~\bibnamefont {Adhikari}}, \ and\
  \bibinfo {author} {\bibfnamefont {M.}~\bibnamefont {Takeoka}},\ }\href
  {\doibase 10.1088/1367-2630/aac11a} {\bibfield  {journal} {\bibinfo
  {journal} {New J. Phys.}\ }\textbf {\bibinfo {volume} {20}},\ \bibinfo
  {pages} {063025} (\bibinfo {year} {2018})}\BibitemShut {NoStop}%
\bibitem [{Note1()}]{Note1}%
  \BibitemOpen
  \bibinfo {note} {A {\protect \em mean function} is a function that defines an
  `average value' of two numbers. Familiar examples of mean functions are the
  arithmetic mean, harmonic mean and geometric mean. A function $\protect
  \mathcal {M}(x,y)$ is a mean function if it can be written as $\protect
  \mathcal {M}(x,y) = y f({x}/{y})$ where $f$ is: (i) monotone increasing, (ii)
  $x f^{-1}(x)=f(x)$ so that $\protect \mathcal {M}$ is symmetric, and (iii)
  $f(1)=1$. See for example Refs.~\cite {petz1996monotone,
  hiai2009riemannian}.}\BibitemShut {Stop}%
\bibitem [{Note2()}]{Note2}%
  \BibitemOpen
  \bibinfo {note} {Numerical optimization was carried out using standard
  techniques and programmes in MATLAB. The coherent and TMSV states were used
  as starting points for the density matrices in the single and two mode case
  respectively. These states were then subject to the operations representing a
  quantum illumination channel and the subsequent error probabilities were
  computed using Eq.~\protect \textup {\hbox {\mathsurround \z@ \protect
  \normalfont \relax \protect \fontsize {10}{12}\protect \selectfont
  \abovedisplayskip 10\p@ plus2\p@ minus5\p@ \belowdisplayskip
  \abovedisplayskip \abovedisplayshortskip \abovedisplayskip
  \belowdisplayshortskip \abovedisplayskip \let \leftmargin \leftmargini
  \parsep 4\p@ plus2\p@ minus\p@ \topsep 8\p@ plus2\p@ minus4\p@ \itemsep 4\p@
  plus2\p@ minus\p@ \leftmargin \leftmargini \parsep 4\p@ plus2\p@ minus\p@
  \topsep 8\p@ plus2\p@ minus4\p@ \itemsep 4\p@ plus2\p@ minus\p@
  (\ignorespaces \ref {kapeqn}\unskip \@@italiccorr )}}. The error probability
  was then minimised over physical density matrices, subject to the energy
  constraint, using fmincon. For this reason, the solutions found might
  correspond to local minima instead of the global minima}\BibitemShut
  {NoStop}%
\bibitem [{\citenamefont {Guha}\ and\ \citenamefont
  {Erkmen}(2009)}]{Guha_Erkmen_PRA_09}%
  \BibitemOpen
  \bibfield  {author} {\bibinfo {author} {\bibfnamefont {S.}~\bibnamefont
  {Guha}}\ and\ \bibinfo {author} {\bibfnamefont {B.~I.}\ \bibnamefont
  {Erkmen}},\ }\href {\doibase 10.1103/PhysRevA.80.052310} {\bibfield
  {journal} {\bibinfo  {journal} {Phys. Rev. A}\ }\textbf {\bibinfo {volume}
  {80}},\ \bibinfo {pages} {052310} (\bibinfo {year} {2009})}\BibitemShut
  {NoStop}%
\bibitem [{\citenamefont {Lee}\ \emph {et~al.}(2020)\citenamefont {Lee},
  \citenamefont {Ihn},\ and\ \citenamefont {Kim}}]{Lee_Ihn_Kim_arxiv_20}%
  \BibitemOpen
  \bibfield  {author} {\bibinfo {author} {\bibfnamefont {S.-Y.}\ \bibnamefont
  {Lee}}, \bibinfo {author} {\bibfnamefont {Y.~S.}\ \bibnamefont {Ihn}}, \ and\
  \bibinfo {author} {\bibfnamefont {Z.}~\bibnamefont {Kim}},\ }\href
  {https://arxiv.org/abs/2004.09234} {\bibfield  {journal} {\bibinfo  {journal}
  {Preprint at arXiv:2004.09234}\ } (\bibinfo {year} {2020})}\BibitemShut
  {NoStop}%
\bibitem [{\citenamefont {De~Palma}\ and\ \citenamefont
  {Borregaard}(2018)}]{DePalma_Borregaard_PRA_18}%
  \BibitemOpen
  \bibfield  {author} {\bibinfo {author} {\bibfnamefont {G.}~\bibnamefont
  {De~Palma}}\ and\ \bibinfo {author} {\bibfnamefont {J.}~\bibnamefont
  {Borregaard}},\ }\href {\doibase 10.1103/PhysRevA.98.012101} {\bibfield
  {journal} {\bibinfo  {journal} {Phys. Rev. A}\ }\textbf {\bibinfo {volume}
  {98}},\ \bibinfo {pages} {012101} (\bibinfo {year} {2018})}\BibitemShut
  {NoStop}%
\bibitem [{\citenamefont {Spedalieri}\ and\ \citenamefont
  {Braunstein}(2014)}]{Spedalieri_Braunstein_PRA_14}%
  \BibitemOpen
  \bibfield  {author} {\bibinfo {author} {\bibfnamefont {G.}~\bibnamefont
  {Spedalieri}}\ and\ \bibinfo {author} {\bibfnamefont {S.~L.}\ \bibnamefont
  {Braunstein}},\ }\href {\doibase 10.1103/PhysRevA.90.052307} {\bibfield
  {journal} {\bibinfo  {journal} {Phys. Rev. A}\ }\textbf {\bibinfo {volume}
  {90}},\ \bibinfo {pages} {052307} (\bibinfo {year} {2014})}\BibitemShut
  {NoStop}%
\bibitem [{\citenamefont {Wilde}\ \emph {et~al.}(2017)\citenamefont {Wilde},
  \citenamefont {Tomamichel}, \citenamefont {Lloyd},\ and\ \citenamefont
  {Berta}}]{Wilde_etal_PRL_17}%
  \BibitemOpen
  \bibfield  {author} {\bibinfo {author} {\bibfnamefont {M.~M.}\ \bibnamefont
  {Wilde}}, \bibinfo {author} {\bibfnamefont {M.}~\bibnamefont {Tomamichel}},
  \bibinfo {author} {\bibfnamefont {S.}~\bibnamefont {Lloyd}}, \ and\ \bibinfo
  {author} {\bibfnamefont {M.}~\bibnamefont {Berta}},\ }\href {\doibase
  10.1103/PhysRevLett.119.120501} {\bibfield  {journal} {\bibinfo  {journal}
  {Phys. Rev. Lett.}\ }\textbf {\bibinfo {volume} {119}},\ \bibinfo {pages}
  {120501} (\bibinfo {year} {2017})}\BibitemShut {NoStop}%
\bibitem [{\citenamefont {Petz}(1996)}]{petz1996monotone}%
  \BibitemOpen
  \bibfield  {author} {\bibinfo {author} {\bibfnamefont {D.}~\bibnamefont
  {Petz}},\ }\href@noop {} {\bibfield  {journal} {\bibinfo  {journal} {Linear
  Algebra and its Applications}\ }\textbf {\bibinfo {volume} {244}},\ \bibinfo
  {pages} {81} (\bibinfo {year} {1996})}\BibitemShut {NoStop}%
\bibitem [{\citenamefont {Hiai}\ and\ \citenamefont
  {Petz}(2009)}]{hiai2009riemannian}%
  \BibitemOpen
  \bibfield  {author} {\bibinfo {author} {\bibfnamefont {F.}~\bibnamefont
  {Hiai}}\ and\ \bibinfo {author} {\bibfnamefont {D.}~\bibnamefont {Petz}},\
  }\href@noop {} {\bibfield  {journal} {\bibinfo  {journal} {Linear Algebra and
  its Applications}\ }\textbf {\bibinfo {volume} {430}},\ \bibinfo {pages}
  {3105} (\bibinfo {year} {2009})}\BibitemShut {NoStop}%
\end{thebibliography}%

\end{document}